\documentclass[traditabstract]{aa}
\usepackage{txfonts,amssymb,graphicx}
\usepackage{natbib}

\begin{document}
\title{An updated estimate of the number of Jupiter-family comets using a simple fading law}
\author{R. Brasser\inst{1,2} \and J.-H. Wang\inst{1}}
\institute{Institute for Astronomy and Astrophysics, Academia Sinica; 11F AS/NTU building, 1 Roosevelt Rd., Sec. 4, Taipei 10617,
Taiwan \and Earth-Life Science Institute, Tokyo Institute of Technology, Meguro, Tokyo 152-8551, Japan
}
\date{}
\abstract{It has long been hypothesised that the Jupiter-family comets (JFCs) come from the scattered disc, an unstable
planetesimal population beyond Neptune. This viewpoint has been widely accepted, but a few issues remain, the most prominent of which
are the total number of visible JFCs with a perihelion distance $q<2.5$~AU and the corresponding number of objects in the Scattered
Disk. In this work we give a robust estimate of the number of visible JFCs with $q<2.5$~AU and diameter $D>2.3$~km based on
recent observational data. This is combined with numerical simulations that use a simple fading law applied to JFCs that come close to
the Sun. For this we numerically evolve thousands of comets from the scattered disc through the realm of the giant planets and keep
track of their number of perihelion passages with perihelion distance $q<2.5$~AU, below which the activity is supposed to increase
considerably. We can simultaneously fit the JFC inclination and semi-major axis distribution accurately with { a delayed power law 
fading function of the form $\Phi_m\propto (M^2+m^2)^{-k/2}$, where $\Phi_m$ is the visibility, $m$ is the number of perihelion 
passages with $q<2.5$~AU, $M$ is an integer constant, and $k$ is the fading index. We best match both the inclination and semi-major 
axis distributions when $k \sim 1.4$, $M=40$, { and the maximum perihelion distance below which the observational data is complete 
is} $q_{\rm m} \sim 2.3$~AU.} From observational data we calculate that a JFC with diameter $D=2.3$~km has a typical total absolute 
magnitude $H_T=10.8$, and the steady-state number of active JFCs with diameter $D>2.3$~km and $q<2.5$~AU is of the order of 300 (but 
with large uncertainties), approximately a factor two higher than earlier estimates. The increased JFC population results in a 
scattered disc population of { 6 billion} objects and decreases the observed Oort cloud to scattered disc population ratio to 13, 
{ virtually the same} as the value of 12 obtained with numerical simulations.}
\keywords{fill in when accepted}
\titlerunning{The Jupiter-family comet population}
\authorrunning{Brasser \& Wang}
\maketitle

\section{Introduction}
The solar system is host to a large population of comets, which tend to be concentrated in three reservoirs: the Oort cloud
\citep{1950BAN....11...91O}, the Kuiper belt and scattered disc \citep{1997Sci...276.1670D}. The third is the source of the so-called
Jupiter-family comets (JFCs), a population of comets whose orbits stay close to the ecliptic
\citep{1997Sci...276.1670D, 2008ApJ...687..714V, 2013Icar..225...40B}. In this study we adhere to the definition of
\cite{1996ASPC..107..173L} which states that a JFC has $T_{\rm J} \in (2,3)$ and $a<7.35$~AU (Period $P<20$~yr). Here $a$ is the
semi-major axis and $T_{\rm J}$ is the Tisserand parameter with respect to Jupiter. In addition, we impose a perihelion
distance $q<5$~AU. By contrast, the Oort cloud is the source of the Halley-type comets (HTCs) \citep{2014A&A...563A.122W},
which are defined as having $T_{\rm J}<2$ and $P<200$~yr \citep{1996ASPC..107..173L}. The long-period comets (LPCs) have $P>200$~yr.\\ 
The origin and dynamical evolution of the JFCs have been intensively investigated. \cite{1997Icar..127...13L} ran many numerical
simulations in which they evolved test particles from the Kuiper belt through the realm of the giant planets until they became
visible JFCs (which are defined as JFCs with perihelion distance $q<2.5$~AU). They found that approximately 30\% of Kuiper belt 
objects became JFCs. The JFC inclination distribution could only be reproduced if the comets faded  or disintegrated after a total 
active lifetime of 12~kyr. This implied that the JFCs spent about 3~kyr, or about 400 returns, while active with $q<2.5$~AU.
\cite{1997Icar..127...13L} conclude that in steady-state there should be approximately 100 JFCs with diameter $D>2$~km and 
$q<2.5$~AU.\\ 
Their work was followed by \cite{1997Sci...276.1670D} who concluded that the scattered disc and not the Kuiper belt had 
to produce the JFCs, a conclusion that was confirmed in subsequent works by \cite{2004MNRAS.350..161E}, \cite{2008ApJ...687..714V}, and
\cite{2013Icar..225...40B}. \cite{1997Sci...276.1670D} also reported that the scattered disc had to contain $6 \times 10^8$ objects
with diameter $D>2$~km, confirmed by \cite{2008ApJ...687..714V}.\\ 
\cite{2002Icar..159..358F} studied the evolution of JFCs in parallel to the above works, and reproduced the median dynamical
lifetime and physical lifetimes reported in \cite{1997Icar..127...13L}. Their work was superseded by that of
\cite{2009Icar..203..140D}, who employed an elaborate splitting and fading mechanism to constrain the JFC population and reproduce the
orbital element distributions. They conclude that the active lifetime is comparable to that found by \cite{1997Sci...276.1670D} and
\cite{2002Icar..159..358F} and that in steady state there are approximately 100 JFCs with diameter $D>2$~km and $q<2.5$~AU.\\
Despite the above successes in reproducing the orbital distribution and number of the JFC population several issues remain.
\cite{2014A&A...563A.122W} successfully reproduced the orbital distribution of the Halley-type comets (HTCs) by employing a simple
fading law. Here we apply the same fading law and techniques to numerical simulations of JFC production, and determine whether this
reproduces their orbital distribution and the increase in total absolute magnitude determined from observations. We use the results of
the fading law to update the number of JFCs with a diameter larger than 2.3~km estimated in \cite{2013Icar..225...40B}. This paper is
organised as follows.\\
In the next section we briefly discuss the observational dataset that we employed. In Section~3 we use the observed absolute
magnitudes and sizes of JFC nuclei to determine their active fraction and the increase in total absolute magnitude from a fully active
comet. Section~4 contains a summary of the numerical simulations that we performed. In Section~5 we discuss the results, Section~6
discusses an important implication of our work. A discussion follows in Section~7 and we draw our conclusions in the last section.

\section{Observational dataset}
In this study we determine whether the orbital distribution of the JFCs can be obtained by using a similar simple fading law that was
applied to the HTCs in \cite{2014A&A...563A.122W}, and to update the number of active JFCs with $q<2.5$~AU. Since the JFCs originate in
scattered disc \citep{1997Sci...276.1670D,2004MNRAS.350..161E,2008ApJ...687..714V,2009Icar..203..140D,2013Icar..225...40B}, we also 
update the inferred number of scattered disc objects (SDOs) by using the total number of active JFCs as a proxy. To compare our 
simulation with the observational data we need to have an observational catalogue of comets that is as complete as possible. Just as in
\cite{2014A&A...563A.122W} we chose to use the JPL Small-Body Search Engine\footnote{http://ssd.jpl.nasa.gov/sbdb\_query.cgi}.\\
In total the catalogue contains 406 JFCs.\\
In addition to their orbital parameters we want to obtain an estimate of the number of JFCs as a function of their size. To do so
requires knowledge of both the total absolute magnitude and the nuclear absolute magnitudes of all the comets or their diameters.
The most complete set of JFC nucleus sizes was reported in \cite{2011MNRAS.414..458S} and the most comprehensive list of total
absolute magnitudes is from \cite{1994P&SS...42..199K}. Using these combined works the total number of JFCs for which both the total
absolute magnitude and diameter are known is 68. We make use of these in Section~5.

\section{Absolute magnitude of JFCs versus LPCs}
\label{JFCvsLPC}
\begin{figure}[]
\resizebox{\hsize}{!}{\includegraphics[angle=-90]{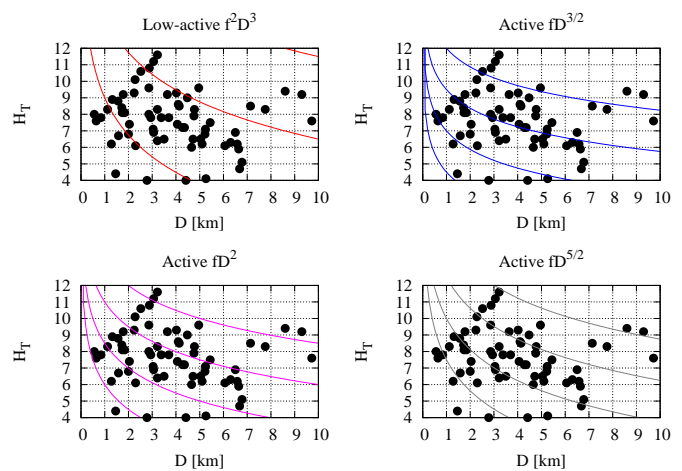}}
\caption{Scatter plot of the diameter versus total absolute magnitude of JFCs. The absolute magnitudes come from
\cite{1994P&SS...42..199K} and the diameters are taken from \cite{2011MNRAS.414..458S}. The lines show the diameter-magnitude relation
for various values of $f$. From bottom to top these are $f=1$, 0.1, 0.01, and 0.001.}
\label{htd}
\end{figure}

Based on the data from \cite{1999A&A...352..327F} and \cite{2006Icar..185..211F}, \cite{2013Icar..225...40B} estimated that a JFC with
$D=2.3$~km has an absolute magnitude of approximately $H_T=9.3$. Here we aim to make a more accurate estimate based on available 
observational data.\\
\cite{1999A&A...352..327F} derive several relations between the total brightness, $B_T$, and diameter of JFCs, $D$, and the
fraction of the surface that shows active outgassing, $f$. If the magnitude limit of the coma is set by its fading into the sky
background - applicable to more distant and smaller comet nuclei and low-active comets - they obtain $B_T \propto f^2D^3$. However, for
comets whose coma is larger than the aperture of the telescope $B_T \propto fD^{3/2}$, while $B_T \propto fD^{5/2}$ for comets far from
the Sun where the activity decreases substantially. A priori it is difficult to know which relation to choose, but generally it
appears to be $B_T \propto fD^{\delta}$ with $3/2<\delta<5/2$. Figure~\ref{htd} depicts the total absolute magnitude of those JFCs for
which their diameters are known. There are a few outliers with $D>10$~km that were excluded. The absolute magnitudes were taken from
\cite{1994P&SS...42..199K} and the diameters are from \cite{2011MNRAS.414..458S}. The lines show the diameter-magnitude relation
for various values of $f$. From bottom to top these are 1, 0.1, 0.01 and 0.001. From the figure and the available data typically $f
\sim 0.01$ when $B_T \propto fD^\delta$, and $f\sim 0.2$ when $B_T \propto f^2D^3$. Are these $D-H_T$ relations and values of $f$ for
JFCs consistent with observational data?\\
\cite{2005ASSL..328.....F} states that for a JFC the typical value of $f$ is around 0.02. To verify this we computed $\log f$ for all
the comets in Fig.~\ref{htd}. This was done as follows. For the remainder of the paper we assume that $B_T \propto fD^2$ and thus
\begin{equation}
H_T = \mathcal{C} - 5\log D -2.5 \log f,
\label{eq:hvsd}
\end{equation}
with $\mathcal{C}$ being a constant. In theory a completely inactive comet has $f=0$, but then equation (\ref{eq:hvsd}) is no longer 
valid. In practice a minimum value of $f$ is attained when $H_T$ is equal to the absolute magnitude of the bare nucleus, which occurs 
typically when $f_{\rm min} \sim 10^{-4}$. { We need to calibrate the constant $\mathcal{C}$, but since the active fraction of JFCs is 
unknown we need to rely on a population of comets where the active fraction is known. These are the LPCs.}\\
From \cite{2011MNRAS.416..767S} we have for the LPCs $H_T = 9.3 - 7.7\log D$ { and  when $D=2.3$~km $H_T=6.5$.} Equating this to 
equation (\ref{eq:hvsd}) and setting $f=1$ we find that $\mathcal{C}=8.31$. Now we can compute $f$ as a function of $H_T$ and $D$ for 
the JFCs whose total absolute magnitude and diameter are known. We find $\log f$ follows a Gaussian distribution with $\langle \log f 
\rangle = -1.73$ { and standard deviation $\sigma = 0.83$}. This reinforces the notion that the JFCs are much less active than 
LPCs of the same size and also invalidates the relation $B_T \propto f^2D^3$ for these comets. The change in $f$ needs to be converted 
into a change in total absolute magnitude, $\Delta H_T$.\\
With each perihelion passage with $q<2.5$~AU the comet loses mass. One may follow di Sisto et al. (2009) to compute how much mass is
lost per perihelion passage, but the end result is the same: the comet fades by reducing the diameter through sublimation and the
active fraction through the formation of insulating layers \citep{1990A&A...237..524R,1991AJ....102.1446R}. The reduction in diameter
and active surface fraction yields a change in absolute magnitude $\Delta H_T = -5\log(D_i/D_f) - 2.5\log(f_i/f_f)$, where subscript i
stands for initial values and subscript f for final values. When one considers a typical mass loss rate of 40~g~cm$^{-2}$ per
perihelion passage \citep{2005ASSL..328.....F}, which occurs when $q \sim2$~AU, then after a few hundred passages for
small comets $\log(D_f/D_i) \sim -0.1$ and, assuming the active fraction stays constant, $\Delta H_T \sim 0.5$.\\ 
However, the active fraction does not stay constant but decreases as well, most likely in accordance with the crust building scenario.
From Fig.~\ref{htd} many observed JFCs have $f \sim 0.01$--$0.1$. Assuming that $f_i \sim 1$ like the LPCs and that after
substantial orbital evolution $f_f$ is in this range, we have $\Delta H_T = 2.5$--$5$ magnitudes. This range of increased
absolute magnitude is consistent with earlier estimates by \cite{1978M&P....18..343W} but lower than those of 
\cite{1999A&A...352..327F} and \cite{2006Icar..185..211F}. Generally, the increase in total absolute magnitude is closer to
the upper end than the lower one. In other words, the greatest increase in the absolute magnitude, i.e. the fading, is caused by the
active fraction of the comet decreasing. A caveat could exist in assuming $f_i \sim 1$, but since we observe some JFCs with a very high
active fraction, these are most likely young comets \citep{1991AJ....102.1446R} and thus this assumption appears justified.\\
In conclusion, a typical JFC is about $\Delta H_T = -2.5\times -1.73 = 4.3$ magnitudes fainter than an LPC of the same size. This is
much fainter than what was used in \cite{2013Icar..225...40B}. We now need to find a fading law that fits the semi-major axis and
inclination distributions of the JFCs, and simultaneously matches the observed amount of fading. The methodology is discussed in the
next sections.

\section {Numerical simulation and initial conditions}
The numerical simulations and initial conditions for this study are described in detail in \cite{2013Icar..225...40B}. In that work
they modelled the formation of the Oort cloud and scattered disc during an episode of giant planet instability. They used the giant 
planet evolution of \cite{2008Icar..196..258L} for the phase of giant planet migration and then continued to evolve the system for an 
additional 4~Gyr, stopping and resuming a few times to clone remaining particles for statistical reasons. However, for this study we 
had to rerun the last 500~Myr of the scattered disc with the same code that was used in \cite{2014A&A...563A.122W} to obtain the 
number of perihelion passages with time of any JFCs we might have produced. For these simulations we used SCATR \citep{Kaib2011}. We 
set the boundary between the regions with short and long time step at 300~AU from the Sun \citep{Kaib2011}. Closer than this distance 
the computations are performed in the heliocentric frame with a time step of 0.1~yr. Farther than 300~AU, the calculations are 
performed in the barycentric frame and we increased the time step to 50~yr. Comets were removed once they were farther than 1000~AU 
from the Sun, or if they collided with the Sun or a planet. { The terrestrial planets were not included because they only have a 
minimal effect on the dynamics of the JFCs \citep{2006Icar..182..161L} and they would increase computation time by at least an order 
of magnitude}.\\

To determine how the comets fade we modified SCATR to keep track of the number of perihelion passages of each comet. By fading we
mean that the comets' visibility (or brightness) decreases. This could be caused by actual fading, splitting, or development of an 
insulating crust. \cite{Levison2001} suggest that comets fade the most quickly when their perihelion distance $q<2.5$~AU, which is the 
distance at which water ice begins to sublimate, so we only counted the number of perihelion passages when the perihelion was closer 
than 2.5~AU.\\
We applied a post-processing { fading law, $\Phi_m$, which is a function of the number of perihelion passages, $m$}. Here 
$\Phi_m$ is the remaining visibility function introduced by \cite{Wiegert1999}. A comet that has its $m^{\rm th}$ perihelion passage 
with $q<2.5$~AU will have its remaining visibility be $\Phi_m$ { up to $m=n_{\rm p}$, where $n_{\rm p}$ is the maximum number of 
perihelion passages before dynamical elimination by Jupiter}; in addition $\Phi_1=1$ for the first out-going perihelion passage of 
a comet, Without fading, every comet in our simulation has the same weighting ($\Phi_m=1,\,\, m=1,2,3 \ldots, n_{\rm p}$) in 
constructing the cumulative distribution of semi-major axis or inclination of active comets. When we imposed the fading effect to 
comets, the remaining visibility $\Phi_m$ is considered as a weighting factor. The higher the number of perihelion passages, the less 
each comet contributes to the cumulative inclination or semi-major axis distribution of the active comets because of fading.\\
In order to find out how well our simulations match with the observed JFCs we follow \cite{2014A&A...563A.122W}. We computed the
cumulative inclination and semi-major axis distributions of the observed and simulated comets. { In each case we imposed a maximum 
perihelion distance, $q_{\rm m}$, below which we deem the observational sample to be complete}, and { employed several functional 
forms of $\Phi_m$.} Once these distributions were generated we performed a Kolmogorov-Smirnov (K-S) test \citep{1992nrca.book.....P}, 
which searches for the maximum absolute deviation $d_{\rm max}$ between the observed and simulated cumulative distributions. The K-S 
test assumes that the entries in the distributions are statistically independent. The probability of a match, $P_{\rm d}$, as a 
function of $d_{\rm max}$, can be calculated to determine whether these two populations were drawn from the same parent distribution. 
However, in our simulations, a single comet would be included many times in the final distribution as long as the comet met our 
criteria for being a JFC during each of its perihelion passages. Including its dynamical evolution in this manner would cause many of 
the entries in the final distribution to become statistically dependent and the K-S test would be inapplicable.\\
We solved this problem by applying a Monte Carlo method to perform the K-S test as described in \cite{Levison2006}. Once we have the
inclination and semi-major axis distributions from the simulation, we then randomly selected 10\,000 fictitious samples from the
simulation. Each sample has the same number of data points as JFCs from JPL catalogue. The Monte Carlo K-S probability is then the
fraction of cases that have their $d$ values between the fictitious samples and real JFC samples larger than the $d_{\rm max}$ found
from the real JFC samples and cumulative distributions from simulation.\\
However, before making the fictitious datasets, we need to find the empirical probability density functions of inclination and
semi-major axis from which we then sample the fictitious JFCs. Here we generated these from a normalised histogram. One crucial
point in making the histogram is that we weighed each entry by its remaining visibility. The fictitious comets were then sampled from
the distributions with the von Neumann rejection technique \citep{vonNeumann1950}. This sampling method relies on generating two
uniform random numbers on a grid. An entry is accepted when both numbers fall under the probability density curve.\\

\section{Results}
{ During the course of investigation we have tried many different forms of $\Phi_m$, but it was necessary to meet a few requirements. 
First, the resulting inclination and semi-major axis distributions of the simulated JFCs need to be consistent with the observed 
sample up to a maximum perihelion distance $q<2.5$~AU. Second, the decay needs a fairly short half-life. The half-life 
is the number of revolutions by which time the visibility, or active surface, has dropped by 50\%. This is equivalent to an increase 
in the total absolute magnitude of 0.75. There are several works that indicate how the comets may decrease their visibility.\\
\cite{1991AJ....102.1446R} studied the fading of comets through orbital changes caused by non-gravitational forces. They reported that 
young, active comets build up an insulating layer in 10 to 20 revolutions and that the brightness of these comets decreases by a 
factor of four within the same number of revolutions. This suggests the fading happens rather quickly. Simultaneously, 
\cite{1991ApJ...366..318P} clearly show that comets fade quickly during the first few returns and much more slowly thereafter. This 
confirms that comets should fade substantially during the first 10 to 20 revolutions. However, both studies only considered pristine 
comets where evaporation occurred at a steady rate, while it is known that these same comets build up an insulating layer farther 
from the Sun \citep{2005ASSL..328.....F} which could slow the fading down. Thus we need a fading function with a reasonably slow 
initial decay that speeds up later.} \\
From our simulations we determined $\langle \log n_p \rangle = 2.62 \pm 0.85$. Thus, a JFC typically passes through perihelion with 
$q<2.5$~AU $\langle n_p \rangle =431_{-357}^{+2530}$ times before Jupiter eliminates it. { Assuming no gradual fading,} the total 
active lifetime $\tau_{\rm vJFC} = 3.1^{+21.1}_{-2.6}$~kyr, where we used the median orbital period of 7.3~yr, consistent with earlier 
estimates. Hence after a few hundred revolutions the fading function should match the current typical observed active fraction. We 
observe both new and old JFCs and thus we assume that a new JFC with $m=1$ has the same absolute magnitude as an LPC of the same size, 
i.e. $f_i=1$. That said, we want to post one word of caution.\\
In our simulations a comet is eliminated from the visible region by collision with a planet or the Sun, or by ejection by Jupiter. In 
addition to these, fading may be in itself an end state since, as comets lose matter, they follow a progressive process of 
disintegration, leaving meteoritic matter disseminated through interplanetary space as remains of the parent comet, before ejection by 
Jupiter. Thus, there are additional loss mechanisms apart from planetary ejection. In this regard, the average expected decrease in 
brightness of 4.3 magnitudes before dynamical ejection can only apply to those members of the population that are large enough to 
withstand this fading unscathed until Jupiter can eject them. With this caveat in mind we subsequently explore several functional 
forms of the fading law and their implications.\\

\subsection{Simple power law}
\begin{figure}[ht]
\resizebox{\hsize}{!}{\includegraphics{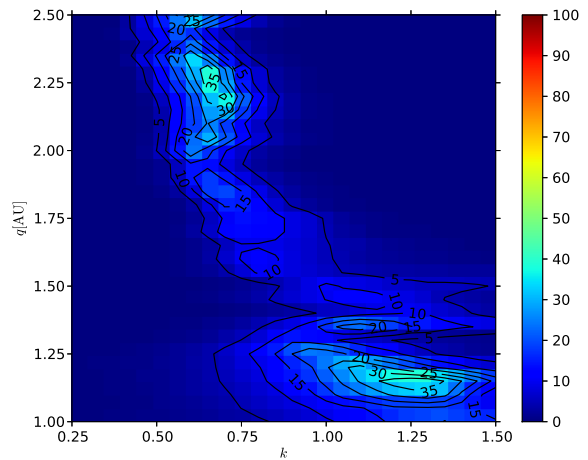}}
\caption{Contour plots of the combined inclination and semi-major axis K-S probability as a function of the fading parameter and 
maximum perihelion distance for the simple power law. The parameters that best fit both distributions are $k=0.65$ and $q_{\rm m} = 
2.25$~AU.}
\label{fig:contour_jfc_ia}
\end{figure}

The results of applying { a fading function of the form $\Phi_m=m^{-k}$} to the simulated JFCs are shown in 
Fig.~\ref{fig:contour_jfc_ia}. Here, { and in subsequent figures}, we plot contours of the product of the K-S probabilities for the 
inclination distribution and semi-major axis distribution as a function of { $q_{\rm m}$ vertically and another variable (in this 
case $k$) horizontally}. In this manner we clearly show where the maxima are for both the inclination and semi-major axis 
distributions simultaneously. We find that the combination of $q_{\rm m}$ and $k$ that best fits both the semi-major axis and 
inclination distribution is $q_{\rm m}=2.25$~AU, $k=0.65$ with a combined K-S probability of 39\% (comprised of 60\% for the 
semi-major axis and 66\% for the inclination). { This result suggests the observational sample is complete up to $q=2.25$~AU.} We 
also note that the probability maxima lie along a band that shifts towards larger $k$ with decreasing $q$. This is expected because 
comets that venture closer tend to evolve faster.\\

There is a second maximum at $q_{\rm m}=1.15$~AU and $k=1.25$. Here the fit for the semi-major axis distribution is very good -
although there are few comets to fit the data to - but the inclination match is poor.\\
What about the total absolute magnitude? Fading causes the total absolute magnitude to increase via $\Delta H_T =-2.5 \log
\Phi_m = 2.5k\log m$. Using $q_{\rm}=2.25$~AU and $k=0.65$, after $\langle n_{\rm p} \rangle$ revolutions the comets have faded by 
$\Delta H_T = 4.3 \pm 1.4$ magnitudes, the same value as was computed in Section~\ref{JFCvsLPC} from observational data. On the other 
hand, for the other maximum with $q_{\rm}=1.15$~AU and $k=1.25$ we need to compute the number of revolutions with $q<1.15$~AU. From 
the perihelion distribution of visible comets the number of JFCs with $q<1.15$~AU is just 12\% of the number with $q<2.5$~AU, from 
which we compute that $\Delta H_T = 5.4$ magnitudes, at least one magnitude higher than for the other maximum. Thus, we report that 
only the first is consistent with the observational data because the second gives too strong a fading. { Unfortunately, even though 
the power law gives an excellent fit to the data for certain values of $q_{\rm m}$ and $k$, the half-life is just $N_{1/2} = 2^{1/k} 
\sim 3$ revolutions, which is much shorter than that advocated by \cite{1991AJ....102.1446R}. For this reason we must discard it in 
favour of a better form.}

\subsection{Constant fading probability}
\begin{figure}
\resizebox{\hsize}{!}{\includegraphics{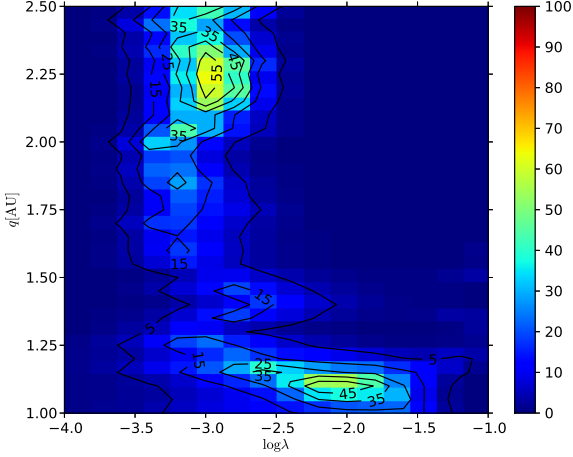}}
\caption{Contour plots of the combined inclination and semi-major axis K-S probability as a function of the fading parameter and
maximum perihelion distance for the constant fading probability case. The $x$-axis is log $\lambda$. The parameters that best fit both 
distributions are $\lambda\sim 0.001$ and $q_{\rm m} = 2.25$~AU.}
\label{fig:contour_jfcql_ia}
\end{figure}

\cite{1994Icar..108..265C} suggest that 1\% of JFCs are destroyed through splitting per perihelion passage. This fading law is 
identical to one suggested by \cite{Wiegert1999}, which is constant fading probability $\Phi_m = (1-\lambda)^{m-1}$, where $\lambda$ 
is the probability of fading (in this case splitting). We proceeded to search for the value of $\lambda$ that fit the orbital 
distribution of the observed JFCs. This is depicted in Fig.~\ref{fig:contour_jfcql_ia}. Once again there are two maxima, one being 
much higher than the other.The best fit, with combined K-S probability 42\% (91\% and 47\% for inclination and semi-major axis) has 
$q_{\rm}=2.25$~AU and $\lambda=0.001$. Unfortunately, this low splitting or fading probability is inconsistent with the dynamical 
simulations because the low probability suggests that the comets' active lifetime is longer than their dynamical lifetime, which is 
untrue. In other words it does not produce the required amount of fading. After $\langle n_{\rm p} \rangle$ revolutions we have 
$\Delta 
H_T = -2.5 (n_{\rm p}-1) \log (1-\lambda) \sim 0.4_{-0.3}^{+2.7}$. Imposing $\lambda=0.01$ { to be consistent with 
\cite{1994Icar..108..265C} does not yield a strong enough fading at the maximum probability near 1.1~AU because the comets only spend 
10\% of the time with $q<1.1$~AU compared to the time they spend with $q<2.5$~AU, so that $\Delta H_T \sim 1.4$.}\\

\subsection{Stretched exponential}
\begin{figure}
\resizebox{\hsize}{!}{\includegraphics{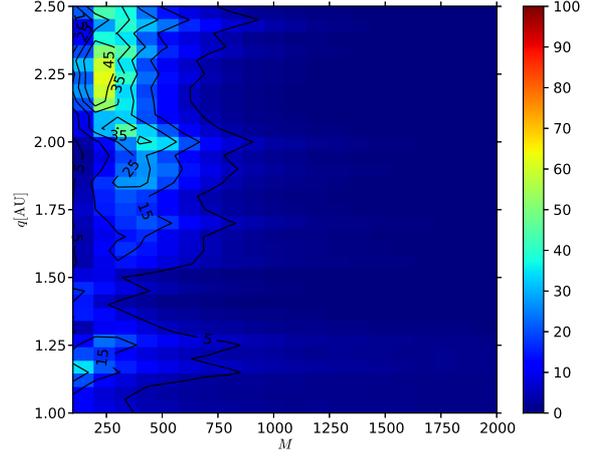}}
\caption{Contour plots of the combined inclination and semi-major axis K-S probability as a function of the fading parameter and 
maximum perihelion distance for the stretched exponential. The stretching parameter is $k=0.2$. The parameters that best fit both 
distributions are $M=200$ and $q_{\rm m} = 2.25$~AU.}
\label{fig:contour_jfcb_ia}
\end{figure}

We next tried a two-parameter fading law of the form $\Phi_m = \exp[-(m/M)^k]$,  which is called the stretched exponential law, and 
$M$ is an integer constant. When the stretching parameter $k<1$ the population suffers infant mortality, while with $k>1$ the 
population suffers from aged mortality. We found that this fading law is able to reproduce the orbital structure at $q_{\rm 
m}=2.25$~AU and several combinations of $M$ and $k$. One example is shown in Fig.~\ref{fig:contour_jfcb_ia} with $k=0.2$, showing 
the combined K-S probability of semi-major axis and inclination as a function of $q_{\rm m}$ and $M$. { In the example, the highest 
combined K-S probability is 63\% (66\% for the semi-major axis and 95\% for the inclination).} All of the combinations of $q_{\rm}$, 
$M$, and $k$ result in a reasonable fading half-life $N_{1/2} = M(\ln 2)^{1/k} \sim 32$ revolutions, but unfortunately the fading at 
high $m$ falls off too slowly to be reconciled with the low active surface fraction of typical JFCs: after $\langle n_{\rm p} \rangle$ 
revolutions we typically have $\Delta H_T =1.09(n_{\rm p}/M)^k \sim 1.2 \pm 0.4$.

\subsection{Delayed power law}
\begin{figure}
\resizebox{\hsize}{!}{\includegraphics{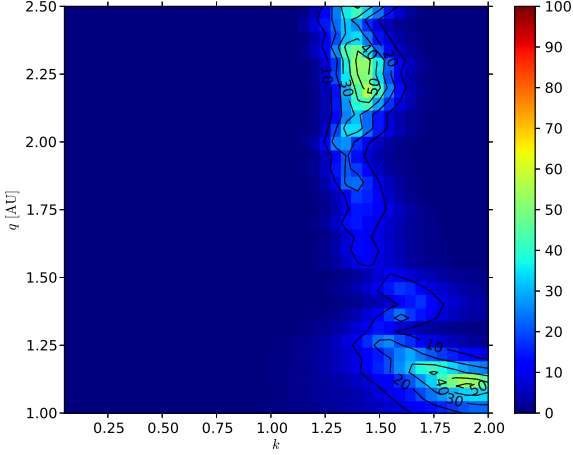}}
\caption{Contour plots of the combined inclination and semi-major axis K-S probability as a function of the fading parameter and 
maximum perihelion distance for the delayed power law. Here $\log M = 1.6$. The parameters that best fit both distributions are 
$k=1.4$ and $q_{\rm m} = 2.30$~AU.}
\label{fig:contour_jfck_ia}
\end{figure}

\begin{figure}[ht]
\resizebox{\hsize}{!}{\includegraphics[]{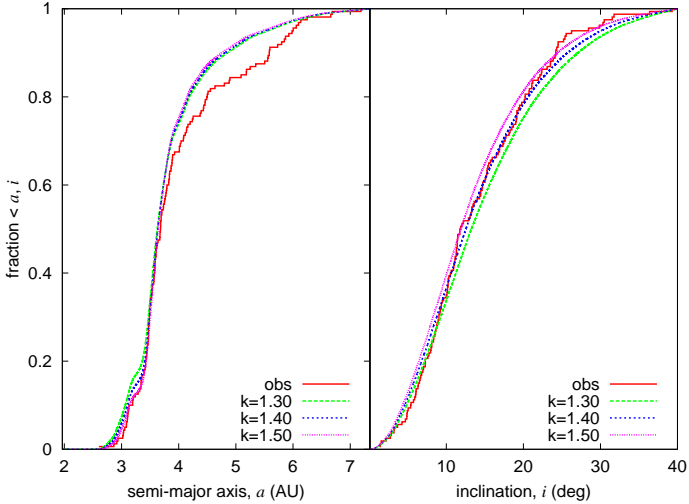}}
\caption{The inclination and semi-major axis distributions from observation and simulation for JFCs that gave the best match from
the delayed power law of Fig.~\ref{fig:contour_jfck_ia}.}
\label{fig:result_jfc_ia}
\end{figure}

The last form we tried was the delayed power law 

\begin{equation}
\Phi_m = \Bigl(\frac{M^2+1}{M^2+m^2}\Bigr)^{k/2},
\end{equation}
where $M$ is an integer constant. The numerator is chosen to make $\Phi_1 = 1$ and at large $m$ the fading proceeds more or less as a 
power law with $m^{-k}$. Once again there are several combinations of $M$ and $k$ that yield good fits. The best fit has $M=40$, the 
probability maximum occurs at $q_{\rm m}=2.3$~AU and $k=1.4$ and is shown in Fig.~\ref{fig:contour_jfck_ia}. The combined maximum 
probability is 53\% (35\% for inclination and 68\% for semi-major axis) and the fits to the cumulative semi-major axis and inclination 
distributions are depicted in Fig.~\ref{fig:result_jfc_ia}. The left panel shows the semi-major axis distribution, the right panel 
depicts the inclination distribution. The median simulated inclination is 12.0$^\circ$ (observed 12.4$^\circ$) and the median 
simulated semi-major axis is 3.74~AU (observed 3.75~AU). The half-life $N_{1/2} \approx M\sqrt{2^{-2/k}-1} \sim 52$ revolutions, 
longer than the 10 to 20 advocated by \cite{1991AJ....102.1446R} and \cite{1991ApJ...366..318P}, but this is likely no problem because 
of the formation of an insulating dust layer. After $\langle n_{\rm p} \rangle$ revolutions the comet has faded by $\Delta H \approx 
2.5k\log (M/n_{\rm p}) \sim 3.6 \pm 2.8$ magnitudes, comparable to the estimation from observations but on the low side. Thus, we 
conclude that this fading law is the most promising because it yields both a high K-S probability, a reasonable match to the total 
active fraction and has a reasonably long half-life. In conclusion, we suggest that the JFCs fade according to this delayed power law. 
In the section below we look at the implications of these results.

\section{Implication: Expected JFC and SDO populations}
We can use the results from the numerical simulations above to constrain the expected number of JFCs larger than a given size,
building on \cite{1997Icar..127...13L} and \cite{2013Icar..225...40B}.\\
In the last few decades many faint JFCs were discovered with dedicated surveys such as LINEAR, Catalina, Spacewatch, Pan-STARRS, that
are not included in \cite{1994P&SS...42..199K}, increasing the average total absolute magnitude of the JFC sample. We can now recompute
the total number of JFCs with $q<2.5$~AU and $D>2.3$~km. From a detailed study of the nuclei and activities of a large sample of JFCs,
\cite{2013Icar..226.1138F} conclude that the JFC population is rather incomplete even for objects with $D>6$~km and $q<2$~AU. Thus, we
need to base our analysis on a sample that is most likely to be complete.\\
We know from the study by \cite{2011MNRAS.416..767S} that an LPC with diameter $D=2.3$~km has $H_T=6.5$, and from observational 
data presented in Section~\ref{JFCvsLPC} a JFC is approximately 4.3 magnitudes fainter than an LPC of the same size. Therefore, a JFC 
with $D=2.3$~km should typically have a total absolute magnitude of $H_T=6.5+4.3=10.8$ rather than $H_T \sim 9$ as was assumed in 
\cite{1997Icar..127...13L} and \cite{2013Icar..225...40B}.\\
\cite{2006Icar..185..211F} state that there are eight JFCs with $H_T<9$ and $q<1.3$~AU. A subsequent search through recently-discovered
JFCs has increased this number to 10. From our simulations and the observational data we find that the fraction of JFCs with $q<1.3$~AU
is 18\% of those with $q<2.5$~AU, and thus the number of JFCs with $q<2.5$~AU and $H_T<9$ is 56. We now need to compute the number of
JFCs with $q<2.5$~AU and $H_T<10.8$ from the absolute magnitude distribution.\\
The cumulative total absolute magnitude distribution of the comets obeys $N(< H_T) \propto 10^{-\alpha_T H_T}$. Since we have imposed
that the total brightness of the comets scales as $B_T \propto fD^2$, it is easy to show that the slope of the total absolute magnitude
distribution, $\alpha_T$, is equal to the slope of the nuclear absolute magnitude distribution, $\alpha$. The latter is related to the
cumulative size-frequency distribution, $N(> D) \propto D^{-\gamma}$, where $\gamma  = 5\alpha$. Even though there is
a lot of scatter in the $D-H_T$ diagram caused by variation in $f$ from one comet to the next \citep{1999A&A...352..327F}, there is a
clear correlation between $D$ and $H_T$ in Fig.~\ref{htd} which must be caused by the underlying size distribution. For JFCs with
diameters between approximately 2~km and 10~km the slope $\gamma \sim 2$ \citep{2004Icar..170..463M,2011MNRAS.414..458S}, corresponding
to $\alpha = 0.4$. The number of JFCs with $q<2.5$~AU and $H_T<10.8$ is then $N_{\rm vJFC}=294_{-235}^{+556}$, about three times
higher than the number reported in \cite{1997Icar..127...13L} and \cite{2009Icar..203..140D}. This is likely to be a lower limit 
because of the aforementioned incompleteness. { However, the fading alters the mean active lifetime, $\tau_{\rm vFJC}$, 
as well. From the output of our simulations and using the delayed power law we computed a weighted mean period of JFCs with $q<2.5$~AU 
of 7.94~yr and a corresponding active lifetime $\tau_{\rm vJFC} =1969^{+7479}_{-1540}$~yr, lower than previous estimates 
\citep{2009Icar..203..140D, 1997Sci...276.1670D,2002Icar..159..358F}.} Following \cite{2013Icar..225...40B} the corresponding number 
of objects in the scattered disc with this updated active lifetime, { taking into account the uncertainties in all relevant 
quantities}, is then $N_{\rm SD}= 5.9^{+2.2}_{-5.1} \times 10^9$. That same work computed an Oort cloud population of $N_{\rm OC} = 
(7.6 \pm 3.3) \times 10^{10}$ for objects with $D>2.3$~km. From our new analysis the Oort cloud to scattered disc population ratio 
turns out to be 13$_{-11}^{+77}$, which { is consistent with} the ratio of 12 $\pm$ 1 from simulations \citep{2013Icar..225...40B}. 
Thus, it is likely that the Oort cloud and scattered disc formed at the same time from the same source, and thus is consistent with a 
formation during the giant planet instability.

\section{Summary and conclusions}
We have performed numerical simulations of the evolution of SDOs until they became visible JFCs. We kept track of their number
of perihelion passages with $q<2.5$~AU and subsequently imposed a form of fading that depended on the number of revolutions. We 
computed the cumulative semi-major axis and inclination distributions and found that these match the observed ones from the JPL 
catalogue when the fading obeys a delayed power law with fading index $k=1.4$, a delay of $M=40$ and the maximum perihelion is 
2.3~AU.\\
From our simulations we find good agreement of the active lifetime of a JFC with earlier works. Our fading law suggests that a typical
JFC has faded by about 3.6 magnitudes before it is eliminated from the solar system by Jupiter or collides with a planet or the Sun, 
ignoring other loss mechanisms such as disintegration and complete evaporation. This increase in $\Delta H_T$ is consistent with 
observational data detailing the active fraction of comets but on the low end. The underlying assumption is that the total brightness 
of the nucleus and coma follows the relation $B_T \propto fD^2$. The above findings imply that the total absolute magnitude of a JFC 
with diameter $D=2.3$~km is 10.8 rather than 9.3 as reported in \cite{1997Icar..127...13L} and \cite{2013Icar..225...40B}. This 
decreased brightness implies there is an increase in the number of JFCs with $D=2.3$~km to approximately 300 rather than the typical 
100 estimated elsewhere. With this updated estimate the number of SDOs reaches { nearly 6} billion and the Oort cloud to scattered 
disc population ratio is estimated to be 13, { virtually the same} as the simulated value of 12 \citep{2013Icar..225...40B}.
\begin{acknowledgements}
We thank Paul Weissman for stimulating discussions that greatly improved this manuscript, Alessandro Morbidelli for pointing out an 
error and Julio Fern\'{a}ndez for a review.
\end{acknowledgements}

\bibliographystyle{apj}

\begin{thebibliography}{16}
\expandafter\ifx\csname natexlab\endcsname\relax\def\natexlab#1{#1}\fi

\bibitem[{{Brasser} \& {Morbidelli}(2013)}]{2013Icar..225...40B} {Brasser}, R. \& {Morbidelli}, A. 2013, \icarus, 225, 40

\bibitem[Chen \& Jewitt(1994)]{1994Icar..108..265C} Chen, J., \& Jewitt, D.\ 1994, \icarus, 108, 265 

\bibitem[{{Duncan} \& {Levison}(1997)}]{1997Sci...276.1670D} {Duncan}, M.~J. \& {Levison}, H.~F. 1997, Science, 276, 1670

\bibitem[Emel'yanenko et al.(2004)]{2004MNRAS.350..161E} Emel'yanenko, V.~V., Asher, D.~J., \& Bailey, M.~E.\ 2004, \mnras, 350, 161 

\bibitem[Fern{\'a}ndez et al.(1999)]{1999A&A...352..327F} Fern{\'a}ndez, J.~A., Tancredi, G., Rickman, H., \& Licandro, J.\ 1999, \aap,
352, 327 

\bibitem[Fern{\'a}ndez et al.(2002)]{2002Icar..159..358F} Fern{\'a}ndez, J.~A., Gallardo, T., \& Brunini, A.\ 2002, \icarus, 159, 358 

\bibitem[Fern{\'a}ndez(2005)]{2005ASSL..328.....F} Fern{\'a}ndez, J.~A.\ 2005, Comets - Nature, Dynamics, Origin, and their 
Cosmogonical Relevance, Astrophysics and Space Science Library, 328, Springer Verlag

\bibitem[Fern{\'a}ndez \& Morbidelli(2006)]{2006Icar..185..211F} Fern{\'a}ndez, J.~A., \& Morbidelli, A.\ 2006, \icarus, 185, 211 

\bibitem[Fern{\'a}ndez et al.(2013)]{2013Icar..226.1138F} Fern{\'a}ndez, Y.~R., Kelley, M.~S., Lamy, P.~L., et al.\ 2013, \icarus, 226,
1138

\bibitem[Hughes(2001)]{2001MNRAS.326..515H} Hughes, D.~W.\ 2001, \mnras, 326, 515 

\bibitem[{{Kaib} {et~al.}(2011){Kaib}, {Quinn}, \& {Brasser}}]{Kaib2011} {Kaib}, N.~A., {Quinn}, T., \& {Brasser}, R. 2011, \aj, 141, 3

\bibitem[Kresak \& Kresakova(1994)]{1994P&SS...42..199K} Kresak, L., \& Kresakova, M.\ 1994, \planss, 42, 199 

\bibitem[{{Levison} \& {Duncan}(1994)}]{Levison1994} {Levison}, H.~F. \& {Duncan}, M.~J. 1994, \icarus, 108, 18

\bibitem[{{Levison}(1996)}]{1996ASPC..107..173L}{Levison}, H.~F. 1996, in Astronomical Society of the Pacific Conference
  Series, Vol. 107, Completing the Inventory of the solar system, ed. T.~{Rettig} \& J.~M. {Hahn}, 173--191

\bibitem[{{Levison} \& {Duncan}(1997)}]{1997Icar..127...13L} {Levison}, H.~F. \& {Duncan}, M.~J. 1997, \icarus, 127, 13

\bibitem[{{Levison} {et~al.}(2001){Levison}, {Dones}, \& {Duncan}}]{Levison2001} {Levison}, H.~F., {Dones}, L., \& {Duncan}, M.~J.
2001, \aj, 121, 2253

\bibitem[{{Levison} {et~al.}(2006)b{Levison}, {Duncan}, {Dones}, \& {Gladman}}]{Levison2006} {Levison}, H.~F., {Duncan}, M.~J., 
{Dones}, L., \& {Gladman}, B.~J. 2006, \icarus, 184, 619

\bibitem[Levison et al.(2006)a]{2006Icar..182..161L} Levison, H.~F., Terrell, D., Wiegert, P.~A., Dones, L., \& Duncan, M.~J.\ 2006, 
\icarus, 182, 161 

\bibitem[Levison et al.(2008)]{2008Icar..196..258L} Levison, H.~F., Morbidelli, A., Van Laerhoven, C., Gomes, R., \& Tsiganis, K.\
2008, \icarus, 196, 258 

\bibitem[{{Marsden} \& {Williams}(2008)}]{Marsden2008} {Marsden}, B.~G. \& {Williams}, G.~V. 2008, Catalogue of Cometary Orbits 2008

\bibitem[Meech et al.(2004)]{2004Icar..170..463M} Meech, K.~J., Hainaut, O.~R., \& Marsden, B.~G.\ 2004, \icarus, 170, 463 

\bibitem[{{von Neumann}(1950)}]{vonNeumann1950} {von Neumann}, J.~V. 1950, {Nat. Bureau Standards 12}, 36-38

\bibitem[{{Oort}(1950)}]{1950BAN....11...91O} {Oort}, J.~H. 1950, \bain, 11, 91

\bibitem[{{Press} {et~al.}(1992){Press}, {Teukolsky}, {Vetterling}, \& {Flannery}}]{1992nrca.book.....P} {Press}, W.~H., {Teukolsky},
S.~A., {Vetterling}, W.~T., \& {Flannery}, B.~P. 1992, {Numerical recipes in C. The art of scientific computing}

\bibitem[Prialnik \& Mekler(1991)]{1991ApJ...366..318P} Prialnik, D., \& Mekler, Y.\ 1991, \apj, 366, 318 

\bibitem[Rickman et al.(1990)]{1990A&A...237..524R} Rickman, H., Fernandez, J.~A., \& Gustafson, B.~A.~S.\ 1990, \aap, 237, 524 

\bibitem[Rickman et al.(1991)]{1991AJ....102.1446R} Rickman, H., Kamel, L., Froeschle, C., \& Festou, M.~C.\ 1991, \aj, 102, 1446 

\bibitem[{{Sosa} \& {Fern{\'a}ndez}(2011)}]{2011MNRAS.416..767S} {Sosa}, A. \& {Fern{\'a}ndez}, J.~A. 2011, \mnras, 416, 767

\bibitem[Di Sisto et al.(2009)]{2009Icar..203..140D} Di Sisto, R.~P., Fern{\'a}ndez, J.~A., \& Brunini, A.\ 2009, \icarus, 203, 140 

\bibitem[Snodgrass et al.(2011)]{2011MNRAS.414..458S} Snodgrass, C., Fitzsimmons, A., Lowry, S.~C., \& Weissman, P.\ 2011, \mnras, 414,
458 

\bibitem[Volk \& Malhotra(2008)]{2008ApJ...687..714V} Volk, K., \& Malhotra, R.\ 2008, \apj, 687, 714 

\bibitem[Wang \& Brasser(2014)]{2014A&A...563A.122W} Wang, J.-H., \& Brasser, R.\ 2014, \aap, 563, A122 

\bibitem[{{Wiegert} \& {Tremaine}(1999)}]{Wiegert1999} {Wiegert}, P. \& {Tremaine}, S. 1999, \icarus, 137, 84

\bibitem[Whipple(1978)]{1978M&P....18..343W} Whipple, F.~L.\ 1978, Moon and Planets, 18, 343
\end{thebibliography}

\end{document}